\documentclass[reprint,amsmath,amssymb,aps,superscriptaddress,twocolumn,prl]{revtex4-1}
\usepackage{graphicx,enumerate,verbatim,bbold}
\usepackage{amsmath,amssymb,amsthm,float,mathrsfs}
\usepackage{dsfont}
\usepackage[colorlinks]{hyperref}
\usepackage[T1]{fontenc}
\usepackage[utf8]{inputenc}
\usepackage{lmodern}
\usepackage{braket}
\usepackage[caption=false]{subfig}
\usepackage{dcolumn}
\usepackage{xcolor}
\usepackage{soul}

\usepackage[normalem]{ulem}
\usepackage{color,soul}

\begin{document}

\title{Topological phases of a dimerized Fermi-Hubbard model for semiconductor nano-lattices}

\author{Nguyen H. Le}

\affiliation{Advanced Technology Institute and Department of Physics, University of Surrey, Guildford GU2 7XH, United Kingdom}

\author{Andrew J. Fisher}
\affiliation{Department of Physics and Astronomy and London Centre for Nanotechnology,
University College London, Gower Street, London WC1E 6BT, United Kingdom}

\author{Neil J. Curson}
\affiliation{Department of  Electronic and Electrical Engineering and London Centre for Nanotechnology,
University College London, Gower Street, London WC1E 6BT, United Kingdom}

\author{Eran Ginossar}
\affiliation{Advanced Technology Institute and Department of Physics, University of Surrey, Guildford GU2 7XH, United Kingdom}

\begin{abstract}
Motivated by recent advances in fabricating artificial lattices in semiconductors and their promise for quantum simulation of topological materials, we study the one-dimensional dimerized Fermi-Hubbard model. We show how the topological phases at half-filling can be characterized by a reduced Zak phase defined based on the reduced density matrix of each spin subsystem. Signatures of bulk-boundary correspondence are observed in the triplon excitation of the bulk and the edge states of uncoupled spins at the boundaries. At quarter-filling we show that owing to the presence of the Hubbard interaction the system can undergo a transition to the topological ground state of the non-interacting Su-Schrieffer-Heeger model with the application of a moderate-strength external magnetic field. We propose a robust experimental realization with a chain of dopant atoms in silicon or gate-defined quantum dots in GaAs where the transition can be probed by measuring the tunneling current through the many-body state of the chain.
\end{abstract} 

\maketitle
\section{Introduction}
Topological phases of matter are among the most exciting developments of modern condensed matter physics \cite{thouless1982,haldane1988,kane2005,bernevig2006}, owing to their rich phenomenology and wide-ranging potential applications from metrology \cite{Pekola2013} to quantum computation \cite{Beenakker2013}. Many experimental platforms have been used to realize these exotic phases of matter such as cold atoms \cite{Atala2013},  photonic lattices \cite{Lu2014,Mukherjee2017} and  engineered solid-state systems including graphene nanoribbons \cite{Rizzo2018,Groning2018}, arrays of carbon monoxide molecules \cite{Gomes2012,Kempkes2019}, and chlorine monolayers \cite{Drost2017} on a copper surface. The band theory of topological insulators (TI) \cite{Bansil2016} based on the independent-electron approximation is well developed and has had many successes. However, in many of the possible experimental platforms for quantum simulation of TI using electrons in solids, such as dopant atoms and gate-defined quantum dots in semiconductors \cite{salfi2016,hensgens2017}, the electron-electron interaction is much stronger than the hopping amplitude of the electrons \cite{Le2017,Dusko2018} and therefore the independent-electron approximation is poor. Topological phases of strongly correlated models form a topic of ongoing active research with intense theoretical  \cite{Rachel2018} and experimental  effort, including recent implementations in cold atoms \cite{Junemann2017} and two-dimensional materials \cite{Nawa2019}. There have been various proposals for the equivalent of the single-particle Berry phase (or Zak phase in one dimension) for the characterization of interacting topological phases, from the magnetic-flux-induced Berry phase \cite{Resta2000,Xiao2010}  to Green's functions \cite{Manmana2012} and entanglement \cite{Jiang2012,Wang2015,Ye2016}. 

Here we discuss one of the simplest one dimensional (1D) models of strongly-correlated TI, the Su-Schrieffer-Heeger-Hubbard (SSHH) model, whose topological properties in various contexts have been investigated using the entanglement entropy \cite{Jiang2012,Wang2015}, the entanglement spectrum \cite{Ye2016}, correlation functions \cite{Barbiero2018}, quench dynamics \cite{Dahan2017} and Berry phase \cite{Guo2011}. The SSHH model describes electrons hopping on a 1D superlattice with staggered hopping amplitudes but uniform local interaction. In this model there exists a charge excitation gap at half-filling due to the on-site interaction (the Mott gap), and another gap at quarter-filling due to dimerization. This opens the possibility of realizing these fillings in experiments, for example by measuring transport while varying the chemical potential and looking for vanishing conductance when the chemical potential lies in the gaps  \cite{Le2017}. For this reason we focus on these two fillings.

We introduce the concept of the reduced many-body Zak phase based on the reduced density matrix of a subsystem, and show that this phase, rather than the normal many-body Zak phase of the full system, should be used for classifying the topological phases at half-filling. This phase jumps from 0 to $\pi$ as the hopping amplitude difference between the even and odd sites changes sign. At half-filling the usual bulk-edge correspondence is manifested in the topological phase transition: the closing and reopening of the eigenenergy gap at the transition point accompanies the appearance of uncorrelated edge states. This is evident in the triplon-excitation spectrum of the dimer chain. In contrast, at quarter-filling the edges remain correlated to the bulk for both signs of the hopping amplitude difference, because of the presence of long-range antiferromagnetic (AFM) order. There is also no gap in the eigenenergy spectrum due to the presence of gapless spin excitations. So the quarter-filled state does not show the characteristics of a TI.  However, we show that  applying an external magnetic field leads to a transition to the topological ground state of the non-interacting Su-Schrieffer-Heeger (SSH) model. Importantly, the strong on-site interaction significantly reduces the critical field strength required for the transition.  Thus our analysis paves the way for the observation of electronic one-dimensional topological insulator states in nanofabricated semiconductor devices.  We propose a device architecture for observing this transition in a one dimensional chain of  dopant atoms or quantum dots. The transition can be probed by measuring the tunneling current through the edges of the chain, which we estimate using a many-body formulation for the conductance of coupled quantum dots \cite{Beenakker1991,Klimeck2008,Chen1994}. 

\section{Results}
\subsection{The Su-Schrieffer-Heeger-Hubbard model}
\begin{figure}[t]
\centering
\includegraphics[width=0.45\textwidth]{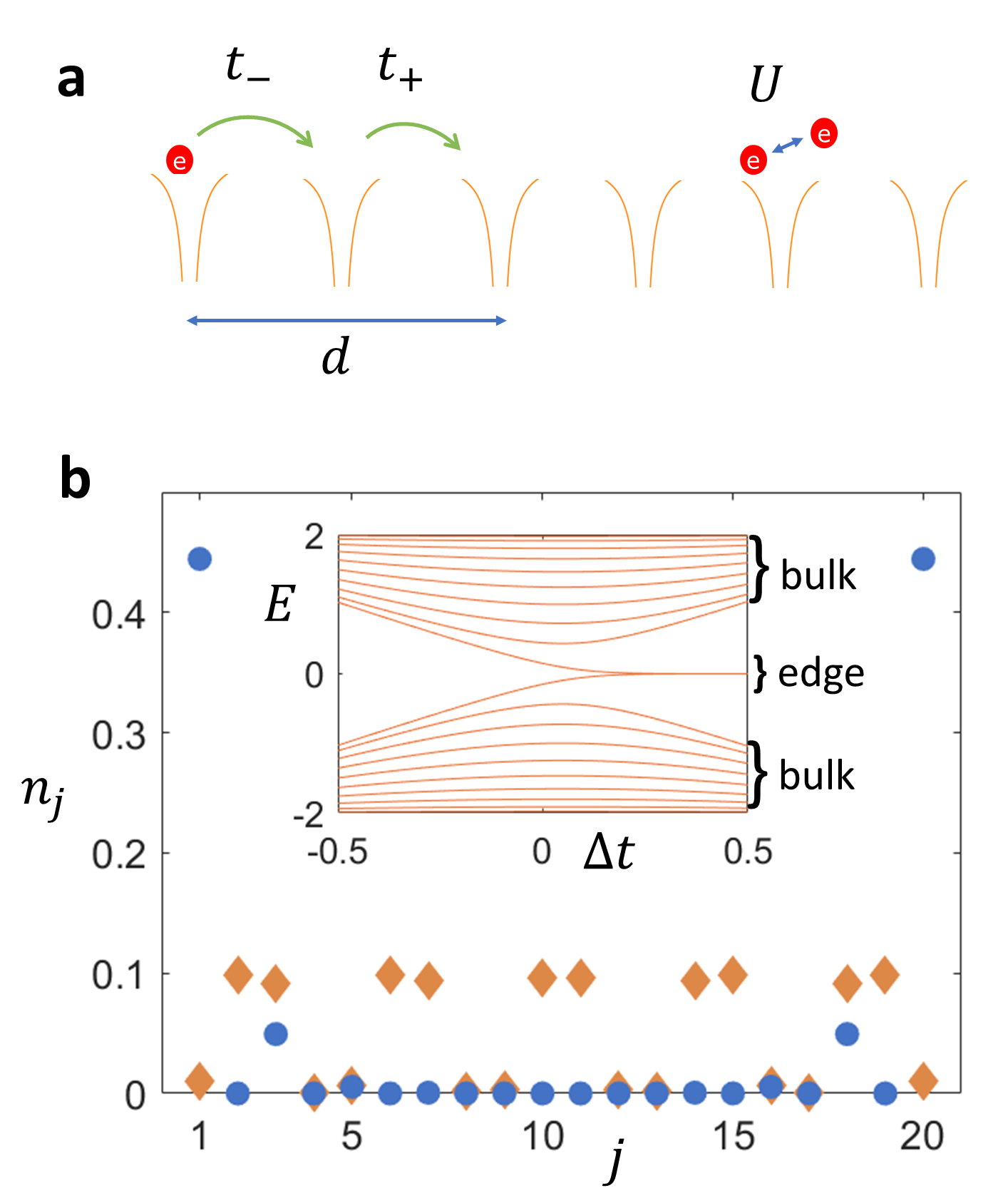}
\caption{\label{fig:SSH} (a) The non-interacting SSH model of particles hopping along a 1D chain of potential wells with alternating hopping amplitudes. (b) Particle distribution of a typical bulk state (diamond), and an edge state (circle) for $\Delta t=0.5$. Inset: Eigenenergies calculated for a chain with $N=20$ sites; the bulk energy gap closes at $\Delta t=0$ and opens again with the  appearance of the mid-gap edge states.}
\end{figure}

The SSHH Hamiltonian is 
\begin{align}\label{eq:SSHH}
H=H_0+V,
\end{align}
where
\[
H_0=\sum_{j,\sigma=\uparrow\downarrow}-\left[1+(-1)^j \Delta t \right]c^{\dag}_{j+1,\sigma}c_{j,\sigma} +\text{h.c.},\]
and 
\[
V=U \sum_j n_{j,\uparrow} n_{j,\downarrow}. 
\]
$H_0$ is the well-known non-interacting  Su-Schrieffer-Heeger (SSH) model \cite{Su1979} of  a particle hopping along a chain with staggered hopping amplitudes, $t_{\pm}=1\pm\Delta t$, as shown in Fig.~\ref{fig:SSH}a, and $V$ is the on-site interaction. Here $c^{\dag}_{j,\sigma}$ denotes the creation operator for the particle at site $j$ and spin $\sigma$. All energies in  this paper are scaled by the mean value of the two hopping amplitudes.

We first describe briefly the topological phases of the SSH model given by $H_0$ \cite{asboth2016,Delplace2011}. For one-dimensional (1D) periodic systems of independent particles, the Berry phase picked up during an adiabatic process when the particle moves across the Bloch states in the Brillouin zone, first discussed by Zak \cite{Zak1989}, is
\begin{equation}
    \phi=i\int_{-\pi/d}^{\pi/d} dk \bra{u_k}\partial_k\ket{u_k},
\end{equation}
where $u_k$ is the periodic part of the Bloch wavefunction, $k$ the crystal momentum, and $d$ the length of the unit cell. 

In a chain with open boundary conditions (OBC) the single-particle eigenstates of this Hamiltonian consist of two distinct types: zero-energy edge states which are localized at the left and right edges, and extended bulk states that avoid the edges. The energy spectrum of the bulk states under periodic boundary conditions (PBC) splits into two bands, $E_{\pm}(k)=\pm\left[2(1+\Delta t^2)+2(1-\Delta t^2)\cos(kd)\right]^{1/2}$. The bulk state wavefunction in PBC has the Bloch form
\begin{equation}
    \psi_k=\sum_j e^{i j k d/2} u_k(j)=\sum_j e^{i j k d/2} e^{i\theta_j(k)} c_j^{\dag}\ket{\varnothing},
\end{equation}
where $\ket{\varnothing}$ is the vacuum, and the phase shift $\theta_j(k)=0$ for the odd sub-lattice and  $\vert E_{\pm}(k) \vert \exp[i\theta_j(k)]=2\cos(kd/2)+2i \Delta t \sin(kd/2)$ for the even sub-lattice. For the SSH model the Zak phase is quantised, more specifically it can only be $0$ or $\pi$ depending on the sign of $\Delta t$ \cite{asboth2016,Delplace2011,Atala2013}. There is a topological phase transition from the trivial phase ($\Delta t<0$), where there is no edge state, to the non-trivial phase ($\Delta t>0$) where the edge states appear. The energy levels of the bulk states form two bands separated by a gap in both phases, and the energy levels of the edge states appear in the middle of the band gap in the non-trivial phase (see Fig.~\ref{fig:SSH}b). The trivial and the non-trivial phases are characterized by the Zak phase of $0$ and $\pi$, respectively \cite{asboth2016}.

\subsection{Charge excitation gap}
\begin{figure*}[t]
\centering
\includegraphics[width=0.95\textwidth]{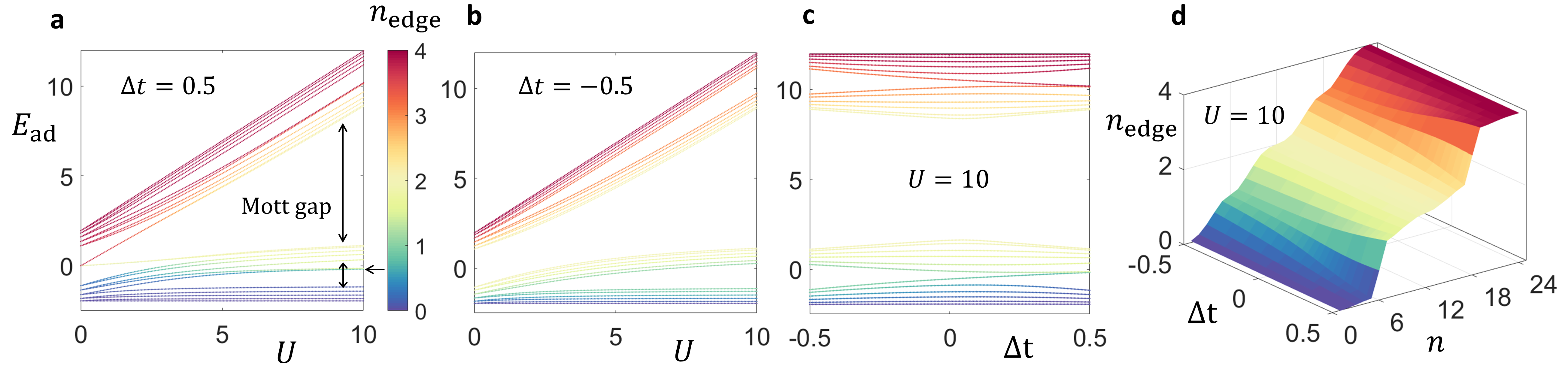}
\caption{\label{fig:Ead} The addition energy spectrum of the Su-Schrieffer-Heeger-Hubbard model as a function of interaction strengths in the (a) non-trivial phase and (b) trivial phase, and (c) as a function of hopping amplitude difference in the strongly correlated limit. To keep track of the edge states the energy levels are colored according to the population at the two ends of the chain ($n_{\text{edge}})$; changes of color thus correspond to occupation of edge states. At strong interaction there are three gaps formed, at half-filling and at quarter/three-quarter-filling. The edge population shows that in the nontrivial phase the mid-gap edge states shift from half-filling at $U=0$ to quarter-filling and three-quarter-fillings at large $U$ [marked by the horizontal arrow in (a)]. The edge population as a function of filling in the strongly correlated limit is shown in (d): In the trivial phase the edge population increases gradually, while in the nontrivial phase the edge population increases sharply at quarter and three-quarter-fillings, indicating the existence of an edge state for the charge excitation at these fillings.}
\end{figure*}

With interaction the single-particle picture is no longer valid, but insight into the topological phases can be gained from looking at the \emph{addition energy spectrum}, also known as the charge excitation spectrum, $E_{\text{ad}}(n)=E_0(n)-E_0(n-1)$ where $E_0(n)$ is the many-body ground energy for filling $n$.  We use exact diagonalization based on the Lanczos algorithm for an open chain with $N=12$ sites to compute the addition energy spectrum and show it in Fig.~\ref{fig:Ead}. \emph{In this paper we set $N=12$ in all the numerical computations for the correlated case}. Features that  survive in the thermodynamic limit are either obvious or stated explicitly.  At $U=0$ the addition energy spectrum reduces to the single particle spectrum of the SSH model with the zero-energy edge state in the middle of the  gap at half-filling. At large interaction the Mott gap forms at half-filling separating the lower and upper Hubbard bands as expected of the Hubbard model. The spectrum has reflection symmetry through the middle point of the Mott gap due to the particle-hole symmetry. Interestingly there are further gaps at one-quarter and three-quarter-fillings, and in the non-trivial phase the edge states of the charge excitation cross to lie in these gaps (see Figs.~\ref{fig:Ead}a and \ref{fig:Ead}d). A more detailed description of these edge states is given below in the discussion of the quarter-filled system. The formation of the quarter-filling gap is due to the combination of the on-site repulsion $U$ and the hopping amplitude difference $\Delta t$, and has been studied previously \cite{Penc94}. Our numerical analysis shows that this gap is approximately $\Delta t$ in the large-U limit. 

The charge gap and the mid-gap edge state in the addition energy spectrum at quarter-filling can be explained analytically in the full dimerization limit where $t_-=0$ and $t_+>0$. The ground state energy level of each Hubbard dimer (coupled by $t_+$) when there is a single particle is the bonding state $E_1=-t_+$, while the ground state energy level of the same dimer with two particles is 
\begin{equation}
E_2=(1/2)\left[U-\sqrt{U^2+16 t_+^2}\right].
\end{equation}

The energy of a particle at the isolated edges is zero. As the particles are added into the chain, they first fill the dimer bonding states as they are lower in energy. When each dimer is filled with one particle we reach the point of quarter-filling. Now if another particle is added to a dimer the energy cost is $\Delta E=E_2-E_1$, while if a particle is added to the edges the energy cost is zero. Hence if $\Delta E>0$ the edges are filled first, otherwise the dimers get filled with two particles until the point of half-filling and only then the edges are filled. Thus the transition of the addition energy level of the edge states from half-filling to quarter-filling happens at the critical interaction $U_c$ such that $\Delta E=0$, or $U_c=3 t_+$. It is obvious from the above discussion that the addition energy gap at quarter-filling is  $\Delta E\approx t_+$ in the strongly correlated limit where $U\gg t_+$. We note that in the general case where neither hopping amplitude is zero the gaps at half-filling and quarter-filling remain in the thermodynamic limit \cite{Penc94}. We provide further evidence by a finite-size scaling analysis of the gap in the Supplementary Material.

\subsection{The reduced many-body Zak phase} 
For correlated systems one can define a many-body Zak phase, first introduced as a measure of macroscopic polarization \cite{Resta1995,Ortiz1994,Grusdt2013}, from the ground state of $H$ satisfying a twisted boundary condition
\begin{equation}\label{eq:tbc}
\Psi_{\kappa}(x_1,...,x_j+L,...,x_N)=e^{i\kappa L}\Psi_{\kappa}(x_1,...,x_j,...,x_N),
\end{equation}
where $L$ is the total length of the chain. Writing $\Psi_{\kappa}=e^{i \kappa\sum_{j=1}^{n} x_j}\Phi_{\kappa}$ with $n$ the number of particles, then $\Phi_{\kappa}$ is the ground state of $H(\kappa)=e^{-i\kappa\sum_{j=1}^{n} x_j}H e^{i \kappa\sum_{j=1}^{n} x_j}$ that satisfies the periodic boundary condition in all coordinates $x_j$.  It can be seen as the many-body analog of $u_k$ in the non-interacting case.  $H(\kappa)$ is the Hamiltonian of a ring threaded with the magnetic flux $\kappa L$ \cite{Kohn1964}, and can be obtained from $H$ by the simple replacement $t_j\rightarrow t_je^{-i\kappa(x_{j+1}-x_j)}$. The many-body Zak phase is then defined as the adiabatic phase picked up by the many-body ground state when this magnetic flux is changed by one flux quantum
\begin{equation}
\phi=i\int_{-\pi/L}^{\pi/L} d \kappa \bra{\Phi_{\kappa}}\partial_\kappa\ket{\Phi_{\kappa}}.
\end{equation}
From Eq.~\eqref{eq:tbc} we see that $\Psi_{-\pi/L}=\Psi_{\pi/L}$ since they satisfy the same anti-periodic boundary condition, hence the function $\Phi_{\kappa}$ at the initial and the end points are related by $\Phi_{\pi/L}=W\Phi_{-\pi/L}$ where $W=e^{-i (2\pi/L) \sum_{j=1}^{n} x_j}$. For numerical computation $\kappa$ is discretized in a grid of $M$ points $\kappa_l$ from $-\pi/L$ to $\pi/L$, and it can be shown that $\phi$ is simply the phase of a complex number \cite{Resta2000}
\begin{equation}\label{eq:zakmb}
\phi=\arg(Z)=\arg\left(\prod_{l=1}^{M-1} \braket{\Phi_{\kappa_l}\vert {\Phi_{\kappa_{l+1}}}}\right).
\end{equation}
This phase can be rewritten in terms of the density matrix $\rho(\kappa)=\ket{\Phi_{\kappa}}\bra{\Phi_{\kappa}}$ as
\begin{equation}\label{eq:zakrho}
\phi=\arg\left[\text{tr} \left(W\prod_{l=1}^{M-1} \rho(\kappa_l)\right)\right].
\end{equation}
The SSHH Hamiltonian $H$ has inversion symmetry, thus changing $x_j$ to $-x_j$ maps $H(\kappa)$ to $H(-\kappa)$. This implies that $H(\kappa)=\mathcal{U} H(-\kappa)\mathcal{U}^{\dagger}$ where $\mathcal{U}$ is the unitary operator of inversion. As a result $E(\kappa)=E(-\kappa)$, and $\Phi_{\kappa}=e^{i\alpha_{\kappa}}\mathcal{U}\Phi_{-\kappa}$, where  $\alpha_{\kappa}$ is an arbitrary phase. It follows that $\rho_{\kappa}=\mathcal{U}\rho_{-\kappa}\mathcal{U}^{\dagger}$, and with a grid centered around $\kappa=0$ such that $\kappa_l=-\kappa_{M-l}$ we have $Z^{*}=\text{tr}\left(\prod_{l=M-1}^{1} \rho(\kappa_l)W^{\dagger}\right)=\text{tr}\left(\prod_{l=1}^{M-1} \rho(\kappa_l)\mathcal{U}^{\dagger}W^{\dagger}\mathcal{U}\right)$. As the inversion operation transforms $x_j$ to $-x_j$ we have $\mathcal{U}^{\dagger}W^{\dagger}\mathcal{U}=W$ and thus $Z^*=Z$, or $Z$ is real, meaning the many-body Zak phase must be either $0$ or $\pi$ depending on whether $Z$ is positive or negative.

The total particle number $n$ is a good quantum number for the eigenstates of the SSHH Hamiltonian. We first study the half-filled spinful case ($n=N$). The ground state $\Phi_{\kappa}$ of $H(\kappa)$ is computed with PBC and the Zak phase is obtained using the discrete formula of Eq.~\eqref{eq:zakmb}. We carry out the computation for $-0.5 \leq \Delta t \leq 0.5$ to see if there is a topological phase change when $\Delta t$ changes sign as in the non-interacting SSH model, and for $0\leq U \leq 10$ to study the effect of interaction. The phase is found to be $0$ for all values of $U$ and $\Delta t$, so it does not reveal any phase transition for either weak or strong interaction. This result is expected at zero interaction, as we then have two copies of the SSH model, one with spin up and the other with spin down. For each copy the Zak phase changes from $0$ to $\pi$, and it is straightforward that the Zak phase of the joint state (given by a Slater determinant) is the sum of the individual phases, hence its values are $0$ and $2\pi$, which are equal since the phase is defined only modulo 2$\pi$. Our result shows that adding interaction does not alter the Zak phase.

For revealing the topological phase transition, we introduce the \emph{reduced Zak phase} of a subsystem within a larger correlated system. It can be obtained from the reduced density matrix of that subsystem. The definition is a generalization of the discrete formula given in Eq.~\eqref{eq:zakrho}, with the density matrix of the total system replaced by the reduced density matrix of the subsystem, which we denote by A:
\begin{equation}\label{eq:reducezak}
\phi_A=\arg(Z_A)=\arg\left[\text{tr} \left(W_A\prod_{l=1}^{M-1} \rho_A(\kappa_l)\right)\right],
\end{equation}
where $\rho_A$ is the reduced density matrix obtained by the partial trace over the complement, $\rho_A(\kappa)=\text{tr}_{\bar{A}}\left(\rho_{\kappa}\right)$, and $W_A=e^{-i (2\pi/L) \sum_{j\in A} x_j}$. For the specific case of the half-filled SSHH model we look at the reduced Zak phases of the spin-up and spin-down subsystem. By symmetry the two phases are equal. We find that both change from $0$ to $\pi$ when $\Delta t$ changes sign for \emph{both weak and strong interaction}, i.e., for all interaction strength in the interval [0, 10].
\begin{figure*}[t]
\centering
\includegraphics[width=0.95\textwidth]{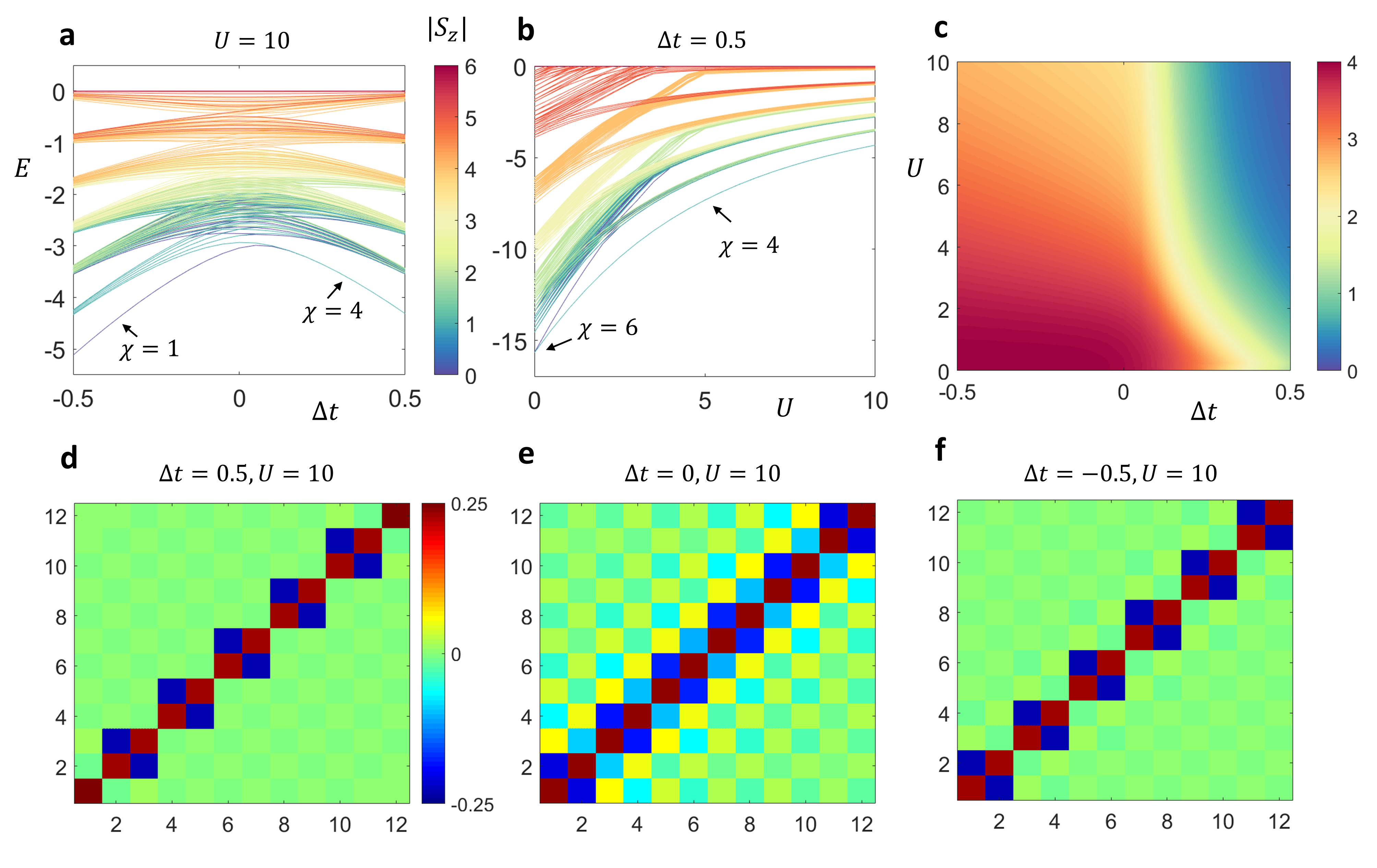}
\caption{\label{fig:hf} Properties of a half-filled SSHH model with OBC. (a) Eigenenergy spectrum in the strongly-correlated regime (U=10), colored according to the spin  $\vert S_z\vert$ of the eigenstates (states with $\pm S_z$ have the same energy). For each value of $S_z$ up to 60 lowest energies are shown. The ground state degeneracy $\chi$ changes from one to four at the topological phase transition due to the four degenerate edge states.   (b) Eigenenergy spectrum in the nontrivial phase ($\Delta t=0.5$) as a function of the on-site interaction. The ground state degeneracy reduces from $\chi=6$ in the noninteracting regime to $\chi=4$ in the interacting regime due to the raising in energy of two ionic states where the edge is double occupied (see text). (c) The entanglement entropy between the first subsystem consisting of the two ends and the second one consisting of the remaining sites in the middle of the chain. (d) Spin correlation $\braket{S_{z,j} S_{z,k}}$  in the nontrivial phase. The abscissa and ordinate are the sites $j$ and $k$. The bulk dimers have perfect AFM correlation but the edges are uncorrelated with the rest; (e) same as (d) for $\Delta t=0$ showing long range AFM correlation; and (f) same as (e) for $\Delta t=-0.5$.}
\end{figure*}

\subsection{Bulk-edge correspondence and triplon excitations}
One of the most interesting aspect of topological band insulators is bulk-edge correspondence: The change of the Zak phase, which is a property of the eigenstates of the bulk, is accompanied by the appearance of localised edge states, and signalled by the closing and reopening of the bulk band gap in the excitation energy spectrum, as demonstrated for the non-interacting SSH model in Fig.~\ref{fig:SSH}. To investigate bulk edge correspondence in the strongly-correlated case at half-filling we compute the excitation energy spectrum of an open chain for $U=10$ in Fig.~\ref{fig:hf}. In our calculations we classify states according to the spin projection, $S_z=(1/2)\sum_{j} (n_{j,\uparrow}-n_{j,\downarrow})$, which commutes with the SSHH Hamiltonian. Owing to the symmetry between spin up and spin down the states with $S_z$ and $-S_z$ have the same energy; the total spin $S$ is also a good quantum number.  In the trivial phase $(\Delta t<0)$, the ground state is non-degenerate and has $S_z=0$ and $S=0$. The energy gap closes at $\Delta t=0$, at which point the lowest triplet exciation ($S=1$, comprising the second lowest energy level with $S_z =0$ and the two lowest energy levels with $S_z=\pm 1$) come down and become degenerate ground states in the non-trivial phase ($\Delta t>0$). Thus the ground state degeneracy $\chi$ changes to $4$ across the topological phase transition.

When $U\gg t_\pm$  the half-filled SSHH model is well approximated by the Heisenberg model of localized spin-1/2 particles  with alternating exchange interaction $J_{\pm}=4t_{\pm}^2/U$ \cite{Essler2010}. For $\Delta t$ far from zero, we find that, in the ground state, each dimer consisting of two sites coupled by the stronger exchange interaction, $J_{>}$, is in a singlet state and is uncorrelated from the other dimers. The excitation gap between the adjacent bands of states in Fig.~\ref{fig:hf}a is due to the excitation to the triplet states of these dimers. This triplon excitation gap is hence approximately $J_{>}$, and the $n$-th excited band has $n$ triplons. 

In the non-trivial phase ($\Delta t>0$) the two edge spins at the two ends of the chain are decoupled from the dimers, and they can be in either a singlet state  
$\ket{S_0}=\left(\ket{\uparrow \downarrow}-\ket{\downarrow \uparrow}\right)/\sqrt{2}$,
or one of the three triplet states $
\ket{T_0}=\left(\ket{\uparrow \downarrow}+\ket{\downarrow \uparrow}\right)/\sqrt{2}$, 
$\ket{T_{+}}=\ket{\uparrow \uparrow}$,
$\ket{T_{-}}=\ket{\downarrow \downarrow}$, all having zero energy. This explains the four-fold degeneracy of the ground state in the non-trivial phase. More specifically, at $\Delta = 0.5$ and $U=10$, our numerical calculation shows that the reduced density matrices of the two edges in the four ground states have around 99\% overlap with the singlet and the triplet states, that is, $\bra{\phi_j}\rho^{\text{edges}}_j\ket{\phi_j}\approx 0.99$ for $j=1,\dots,4$ where $\ket{\phi_j}$ are $\ket{S_0},\ket{T_0}$ and $\ket{T_{\pm}}$. 

The reduced density matrix of each strongly coupled dimer in the bulk at both $\Delta t =\pm 0.5$ satisfies $\bra{S_0}\rho^{\text{dimer}}\ket{S_0}\approx 0.92$. Despite the significant nonvanishing value of the coupling $J_{<}$, the dimers are almost totally uncorrelated with each other and the edges. This is due to monogamy of entanglement: Since each spin in a dimer is maximally entangled with its partner in the singlet state, it must be unentangled from any other spin \cite{koashi2004}.
\begin{figure*}[t]
\centering
\includegraphics[width=0.95\textwidth]{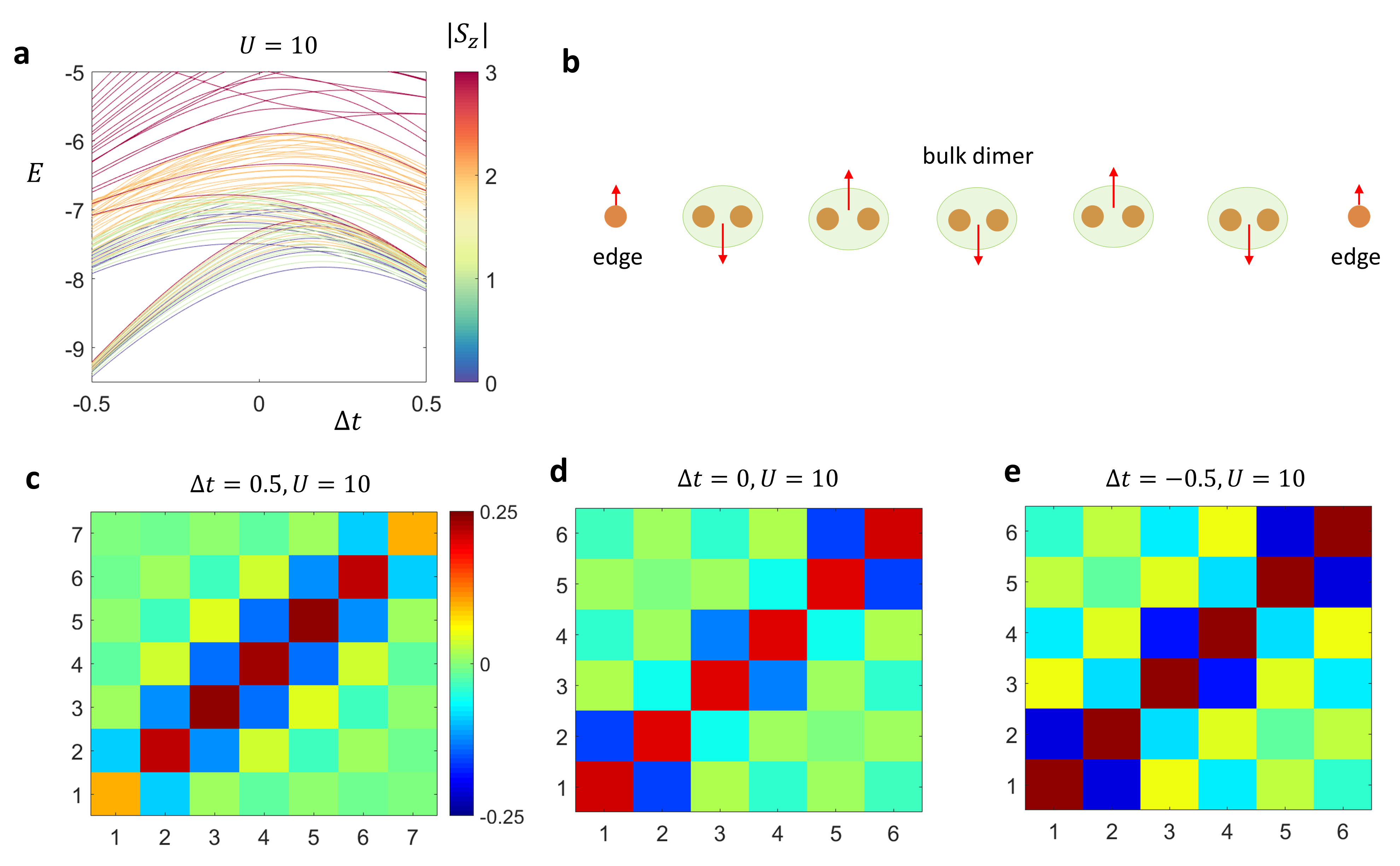}
\caption{\label{fig:qf} Properties of a quarter filled SSHH model with OBC. (a) Eigenenergy spectrum in the strongly-correlated regime ($U=10$), colored according to the spin $\vert S_z\vert$ of the eigenstates. For each value of $S_z$ the 40 lowest energies are shown. (b) An illustration of the particle occupation at each site and the long range AFM correlation between the electron's spins in the non-trivial phase . Each dimer in the middle of the chain is occupied by an electron, and the two edges are shared by one electron.  (c) Spin correlation $\braket{S_{z,j} S_{z,k}}$ in the nontrivial phase ($\Delta t=0.5$). The abscissa and ordinate are the "effective sites" $j$ and $k$ defined as illustrated in (b): The first and the last value refers to the two edges, while each value in between refers to each bulk dimer which is occupied by a single electron;  (d) same as (c) for $\Delta t=0$; and (e) same as (d) for $\Delta t=-0.5$. The long range AFM correlation persists for all values of $\Delta t$.}
\end{figure*}

The decoupled dimers picture is further confirmed by the spin correlation $\braket{S_{z,j} S_{z,k}}$ in the ground state with $S_z=0$, as shown in Fig.~\ref{fig:hf}d. In the nontrivial phase the two spins in the dimer are perfectly anti-correlated with each other, but uncorrelated with other spins, and the edge spins at the two ends are free. The spin correlation in the trivial phase is the same but without the two free ends (see Fig.~\ref{fig:hf}f), since there is no weakly coupled edges in this case. A long range AFM order develops near $\Delta t=0$ as expected for a 1D Hubbard model \cite{Lieb1968} (see Fig.~\ref{fig:hf}e).

In the non-interacting regime the edge states are identified by the localization of a single-particle state at the edges, while in the strongly-correlated regime we have an effective spin model and the edge states are identified as uncorrelated spins. In order to identify the edge states at arbitrary interaction we calculate the von Neumann entropy of the entanglement\cite{Gu2004} between the two ends of the chain and the rest in Fig.~\ref{fig:hf}c. The formation of the edge states is indicated by the sharp drop in entanglement since these states are uncorrelated from the rest. Figure~\ref{fig:hf}c shows a clear transition line where the edge states are formed, and the transition to the edge phase is more abrupt at large $U$. At $U=0$ the transition point is not at $\Delta t=0$ as expected for the SSH model owing to finite-size effects; we expect the transition point would become closer to $\Delta t=0$ with increasing system size (see Supplementary Material).

Figure~\ref{fig:hf}b shows the eigenenergy spectrum in the non-trivial phase as the on-site interaction changes from weak to strong. At $U=0$ the ground state degeneracy is $\chi=6$ since in addition to the one singlet state and three triplet sates the edges have two more degenerate ionic states where two electrons occupy the same edge, i.e. $\ket{\uparrow\downarrow,\varnothing}$ and $\ket{\varnothing,\uparrow\downarrow}$. For $U>0$ the energy of these ionic states is lifted due to the on-site repulsion, thus reducing the ground state degeneracy to $\chi=4$.

\subsection{Persistent long range AFM order at quarter-filling}

Unlike the half-filled case, the quarter-filled system does not exhibit conventional bulk-edge correspondence; this is clear from the lack of a gap in the eigenenergy spectrum (see Fig~\ref{fig:qf}a). Recall that while there exists a gap in the charge excitation spectrum at quarter-filling discussed earlier, the absence of an eigenenergy gap is due to gapless spin excitations \cite{Essler2010}. The picture of decoupled dimers each in a singlet state no longer applies: deep in both phases ($\Delta t$  far from 0) each dimer is occupied by roughly a single electron, and in the nontrivial phase the two ends are shared by one electron (see Fig.~\ref{fig:qf}b for illustration).  In contrast to the isolated edge spins at half-filling, we find that the edges are strongly correlated with the dimers in the bulk through the spin degree of freedom, and there is long-range AFM order in both trivial and non-trivial phases. To show this we again plot the spin correlation   $\braket{S_{z,j} S_{z,k}}$, however now with a different definition for the ``sites'' $j$ and $k$: in the non-trivial phase we denote the first value, $j,k=1$, and the last value, $j,k=N/2+1$, for the left and right edges, and each value in between, $j,k=2,...,N/2$, is assigned to a dimer in the bulk. A similar definition of effective sites is used in the trivial phase, but without the edges. The spin correlation between the edges and the bulk dimers in the non-trivial phase is shown in Fig.~\ref{fig:qf}c; long-range AFM order is visible. The spin correlation in the trivial phase is exactly the same but with the two edges removed as evident in Fig.~\ref{fig:qf}e. Long range AFM order persists in both phases and also at the point of the phase transition (see Fig.~\ref{fig:qf}d). This persistent long range AFM order is the reason why the entanglement between the two edges and the bulk shows no clear drop when $\Delta t$ changes sign (see Fig.~\ref{fig:EB}b and the discussion below). 
\section{Discussion}
\subsection{Magnetic field induced transition to SSH ground state} 
\begin{figure*}[t]
\centering
\includegraphics[width=0.95\textwidth]{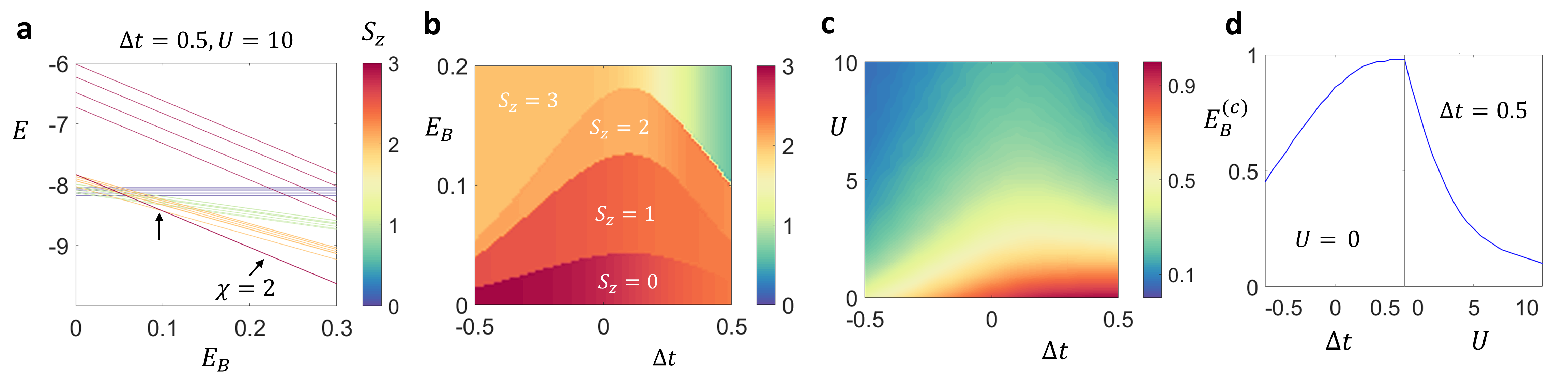}
\caption{\label{fig:EB} Properties of a quarter-filled SSHH model with OBC under an external magnetic field. (a) Eigenenergy spectrum as a function of the field strength in the nontrivial phase. The levels are colored according to the spin projection $S_z$ of the eigenstates. Only states with positive $S_z$ are considered as the ones with negative $S_z$ rise in energy in a magnetic field, and for each value of $S_z$ the 10 lowest energies are shown. At the critical value $E_B^{(c)}\approx 0.1$ (indicated by the arrow) the ground state becomes maximally ferromagnetic (with all spins aligned along the field axis) and reduces to the non-interacting limit of the SSH model. (b) Entanglement entropy between the edges and the bulk  for various field strength and $\Delta t$. Each contour separates regions where the ground state has different $S_z$ ($S_z=3$ indicates the maximally ferromagnetic ground state as there are 6 particles at quarter-filling). At zero field the entanglement does not drop as $\Delta t$ changes sign due to the long range AFM correlation between the edges and the bulk at quarter-filling discussed above. At high field the system reduces to the SSH model, and the entanglement drops for $\Delta t>0$ due to the formation of the localised edge states, signaling the transition to the non-trivial phase.  (c) Critical field strength at various values of the on-site interaction and hopping amplitude difference. (d) 1D slices of (c) for $U=0$ (left panel) and $\Delta t=0.5$ (right panel) showing the sharp decrease in the critical field strength with increasing on-site interaction. }
\end{figure*}

The absence of an eigenenergy gap and uncorrelated edge states suggests that the SSHH model at quarter-filling is not a TI. However, the ground state can undergo a transition to the topological ground state of the non-interacting SSH model if a magnetic field is applied, resulting in the total Hamiltonian $H_1=H-(E_B/2) \sum_j (n_{j,\uparrow}-n_{j,\downarrow})$ where $E_B=g \mu_B B$ and $\mu_B$ is the Bohr magneton and $g$ the g-factor of electrons in the material. At a critical magnetic field the ground state becomes the maximally ferromagnetic state with all of the electron spins aligned along the field axis (see Fig.~\ref{fig:EB}a), and an energy gap is opened. Since there is no on-site interaction between particles with the same spin, this ground state must be the ground state of the non-interacting SSH model with $N/2$ electrons, and due to Pauli exclusion principle the highest energetic electron must occupy the mid-gap edge state shown in Fig.~\ref{fig:SSH}b. We find that the ground state degeneracy at field strength larger than the critical value is $\chi=2$, agreeing with the fact that there are two degenerate edge states (left and right) in the SSH model.  The transition to the SSH ground state is further confirmed in the entanglement entropy between the edges and the bulk in Fig.~\ref{fig:EB}b. Without the field the entanglement does not drop as $\Delta t$ changes sign since the edges are correlated with the bulk through the persistent long range AFM order at quarter-filling. At field strengths beyond the critical value the entanglement drops sharply for $\Delta t>0$ owing to the formation of the localised edge states in the non-trivial phase of the SSH model.

The magnetic-field-induced transition enables experimental realization of the SSH model in systems with local interactions. It is shown in Fig.~\ref{fig:EB}c and more clearly in Fig.~\ref{fig:EB}d, right panel, that the critical magnetic field  reduces dramatically with increasing on-site interaction, meaning it is easier to realize the SSH model if the local interaction is stronger. For reaching the maximally ferromagnetic state the last spin-down particle needs to be pumped to the next unoccupied single-particle energy level shown in Fig.~\ref{fig:SSH}b. When there is strong on-site interaction this spin-down particle interacts strongly with the other particles with opposite spins, raising the energy, hence it costs less energy to pump this spin-down particle to a higher energy level, leading to a smaller required magnetic field. 

Typical parameter values for dopant atoms in silicon and quantum dots in GaAs are $\bar{t}\sim 1\,$meV, $U/\bar{t}\sim 10$, and assuming a g-factor of 2 our calculation gives a critical magnetic field of $B_c \sim 2$T deep in the non-trivial phase at $\Delta t/\bar{t}=0.5$, which is feasible. For electrons bounded to impurities or a quantum dot in a semiconductor host the g-factor can deviate from the free electron value of 2 \cite{Pryor2006}, and the calculated critical field need to be rescaled. As long as the the deviation is not too large the critical magnetic field remains in the realistic range. One sees from Fig.~\ref{fig:EB}d  that without the strong on-site interaction, at $U=0$ for example, the required magnetic field is around ten times larger and therefore may not be realistically attainable for larger hopping amplitude. This emphasizes the importance of interaction.    

We note that the jump in the critical field as  $\Delta t$ changes sign  in the $U=0$ limit (see Fig.~\ref{fig:EB}d, left panel) is due to the formation of the mid-gap edge state in the non-trivial phase of the SSH model. In the trivial phase the last spin-down particle needs to be pumped from the lowest energy to the highest energy in  the lower bulk band of Fig.~\ref{fig:SSH}b, while in the non-trivial phase it needs to be pumped to the mid-gap edge state, which is higher in energy; hence, a larger magnetic field is necessary.

\subsection{Experimental realization with nanofabricated semiconductor devices} 
\begin{figure*}[t]
\centering
\includegraphics[width=0.95\textwidth]{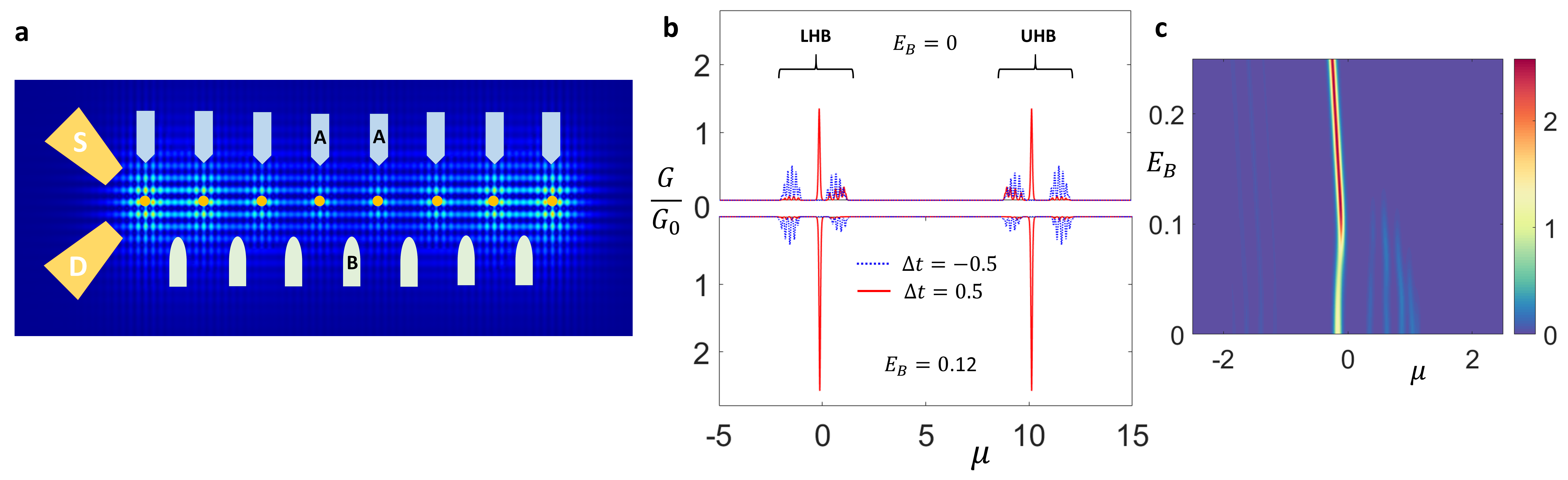}
\caption{\label{fig:exp} Experimental proposal. (a) A device architecture for realizing and probing the topological phase transition of the SSH model. A chain of STM-fabricated dopant atoms, or gate defined quantum dots, is positioned in the centre of a collection of electrodes and gates. The source and the drain leads are used for measuring the current of the electrons tunneling from the source, through the many-body state of the chain, to the drain. Since the source and the drain are close to the edges, the conductance is proportional to the charge excitation density at the edge of the many-body state (see text).  The A side gates are used for tuning the on-site energy, and thus the chemical potential. The B side gates between the sites are used to control the hopping amplitudes. (b) The conductance spectrum at $kT=0.02$ as a function of the chemical potential for $\Delta t=0.5$ (solid red) and $\Delta t=-0.5$ (dotted blue) at zero field (top panel) and at a field strength just above the critical value (bottom panel). The spectrum for both values of $\Delta t$ shows the lower Hubbard band (LHB) and upper Hubbard band (UHB) separated by the Mott gap, and each band is further separated by the charge excitation gap at quarter-filling. For $\Delta t>0$ there is a conductance peak in the middle of this gap due to the tunneling through the edge states of the charge excitation.  The formation of the strongly localised edge state in the SSH model above the critical field strength leads to a sharp increase in the conductance peak at quarter-filling in the bottom panel. (c) Density plot of the conductance around quarter-filling, showing the sharp increase of the peak at the critical field strength. All energies are scaled by the average value of the two hopping amplitudes.}
\end{figure*}

In the last section of the paper we propose a device architecture for realizing the transition to the topological phase of the SSH model described above. With  state-of-the-art fabrication technology it is possible to fabricate 1D chains of dopant atoms in silicon with STM \cite{Zwanenburg2013,Fuechsle2012}, or gate-defined quantum dots in GaAs \cite{hensgens2017}, illustrated in Fig.~\ref{fig:exp}a. Two leads, source and drain, are positioned close to one edge of the chain. Naturally there are potential barriers between these leads and the chain. Electrons from the source can tunnel through the barrier into the many-body state of the chain, and then out to the drain. The A side gates are for tuning the on-site energy, and thus the chemical potential, by applying a voltage. The B side gates between the sites are for controlling the hopping amplitude (also commonly referred to as the ``tunnel coupling" in these systems). A similar device without the source and the drain was fabricated for a chain of 3 quantum dots in Ref.~\cite{hensgens2017}.

We now show how a transport measurement of the proposed device can probe the transition to the SSH ground state at quarter-filling and also the transition between the trivial and nontrivial SSH topological phases. When the tunneling rate $\Gamma$ of the electron from the source/drain to the nearest site is much smaller than the hopping amplitudes between the sites and $kT$, which is typical for dopant atoms and quantum dots, we are in the sequential tunneling regime \cite{Le2017,Lansbergen2010}.  As the chemical potential is varied, each time it matches an addition energy (shown in Fig.~\ref{fig:Ead}), the electron in the lead has enough energy to tunnel into the many-body state of the chain and out to the other lead, resulting in a peak in the tunneling current (see Fig.~\ref{fig:exp}b). Thus the set of peaks in the conductance spectrum maps the addition energy spectrum. The conductance in the linear response regime, applicable when the bias between the source and the drain is much smaller than the hopping amplitudes and $kT$, is computed with the Beenakker's formula \cite{Beenakker1991,Chen1994,Klimeck2008,Le2017}.

For the measurement using the source and drain on the left of Fig.~\ref{fig:exp}a, the conductance peak at filling $n$ is proportional to $G_0 D_n$  where $G_0=e^2\Gamma/(\hbar kT)$ and $D_n=|\braket{\Psi^{(n)}_{0}|c^{\dagger}_{1\uparrow}+c^{\dagger}_{1\downarrow}|\Psi^{(n-1)}_0}|^2$ where $\Psi_0^{(n)}$ is the many-body ground state of the chain at filling $n$. $D_n$ can be interpreted as the charge excitation density at the left edge at filling $n$. Figure \ref{fig:exp}b, top panel, shows the conductance spectrum in the strongly correlated case ($U=10$) for both signs of $\Delta t$, revealing the lower and upper Hubbard bands, separated by the Mott gap, of the addition energy spectrum in Fig.~\ref{fig:Ead}. There is also evidence of the charge gap at quarter-filling in the lower band and three quarter-filling in the upper band. For $\Delta t>0$ there is a sharp conductance peak in the middle of the quarter-filling gap due to the edge state of the charge excitation at this filling, which has a high density of charge excitation at the edges. The lower panel shows the conductance spectrum at an applied field strength just above the critical value of the transition to the SSH model. The mid-gap conductance peaks are much higher since in the SSH model the edge states are much more localised. Observing a sharp increase in the conductance peak at quarter-filling, as shown by our calculation in Fig.~\ref{fig:EB}c, can serve as experimental evidence of the transition to the SSH model of topological insulators. And the appearance of this peak as $\Delta t$ changes sign from negative to positive can be a probe of the topological phase transition.

As an example we consider parameter values that are typical of phosphorous donors in silicon: $\Gamma=0.001$ meV, $\bar{t}=4$ meV and $U=40$ meV. The conductance peak at quarter-filling is then of the order of $10^{-6}$ S at 1K, leading to a tunneling current of 0.1 nA at 0.1 mV bias, which is large enough for detection.

In the above, we discuss a relatively long chain of 12 sites so that the lower and upper Hubbard bands in the conductance spectrum appear dense, but finite-size signature of the edge state can be observed with a much smaller number of sites in experiments, as low as $N=4$. For $N=4$ the `bulk' consists of a single dimer in the non-trivial phase and each half-band either side of the quarter-filling gap in fig.~\ref{fig:exp}b has a single peak. It is better to have $N=6$ so that each half-band has two close peaks and hence can be easily distinguished from the edge-state peak in the middle of the gap. These system sizes are feasible with current technology.

\subsection{Robustness against disorders} 

Disorders in the on-site energy and hopping amplitude are unavoidable in real experiments. We investigate the effect of disorders by adding to the SSHH Hamiltonian the on-site energy term $
  H_{\text{on-site}}=\sum_{j,\sigma=\uparrow \downarrow} \epsilon_{j} n_{j,\sigma}$,
where each $\epsilon_{j}$ is chosen uniformly at random from a range $[-\delta E, \delta E]$, and we add to each hopping amplitude, $t_j=1+(-1)^j\Delta t$, a variation chosen uniformly at random from a range $[-\delta t, \delta t]$. To study the robustness of the magnetic-field induced topological phases above the critical field strength we look at the distribution of the addition energy gap at quarter filling, the distribution of the critical magnetic field required, the signature of the edge state in the conductance spectrum, and the distribution of the many-body Zak phase.
\begin{figure*}[t]
\centering
\includegraphics[width=1\textwidth]{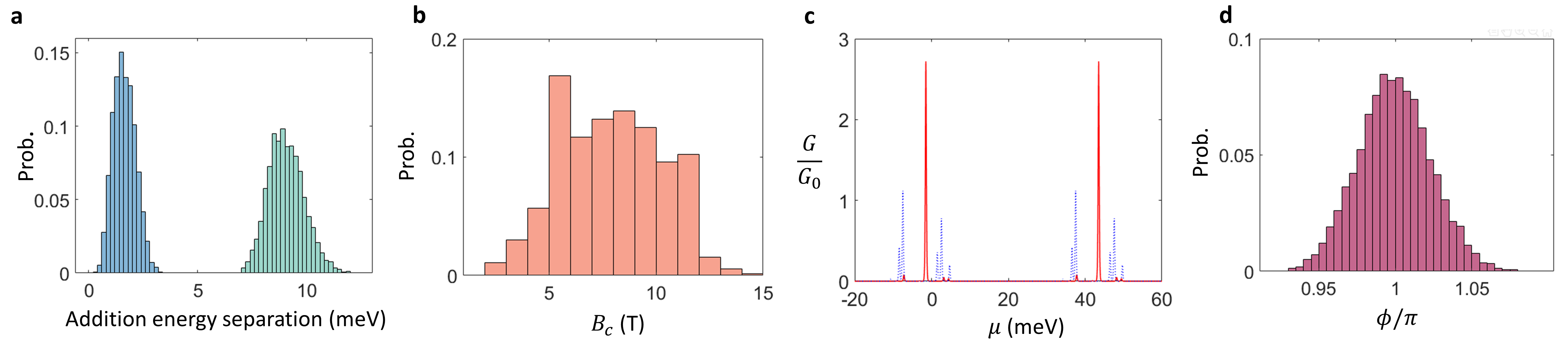}
\caption{\label{fig:qfdis} Robustness of the magnetic-field induced topological phases against disorders for a chain with $N=6$ sites. The mean values of the weak and strong hopping amplitudes are $2$ meV and $6$ meV, and hence $\vert\Delta t\vert$, which is half of the hopping amplitude difference, is 2 meV. The on-site interaction is $U=40$ meV. The maximum variations in the on-site energy and hopping amplitude disorder are both $\vert\Delta t\vert/2$ (see text). The distributions are generated from a sample of 5000 random instances. (a) Probability distributions of the addition energy gap at quarter filling for $B>B_c$ (dark green), and a finite-size addition energy separation within the lower Hubbard band (dark blue), showing that the gap is much larger and distinct from other energy differences arising from finite-size effects. (b) Probability distribution of the critical magnetic field $B_c$ assuming a g-factor of 2. (c) A typical conductance spectrum of the disordered system for $B>B_c$ in the trivial phase (dashed blue) and non-trivial phase (solid red). The appearance of the high rising edge-state peak in the nontrivial phase is clearly visible. (d) Probability distribution of the many-body Zak phase for $B>B_c$, which shows only a small deviation from the ideal value of $\pi$. }
\end{figure*}

We find that, in the presence of both on-site energy and hopping amplitude disorder, the topological phases are robust as long as $\delta t+\delta E< \vert\Delta t\vert$. This is expected since the gap at quarter filling in the fully spin-polarized regime is $\Delta t$, and the above condition makes sure that the gap is not closed by the disorder. Also, $\delta t<\vert \Delta t \vert$ means that the weak-strong order between the odd and even hopping amplitudes is preserved, that is, the couplings within the bulk dimers are always stronger than those between them and with the edges. We show in Fig.~\ref{fig:qfdis} the numerical evidence for robustness when $\Delta t=0.5$ and $\delta t=\delta E= 0.5 \Delta t$. Note that only in this figure we choose to use absolute unit in meV for a more direct connection with experiments.  

Figure \ref{fig:qfdis}a shows the distribution of the addition energy gap at quarter filling, which is $E_{\text{ad}}(N/2+1)-E_{\text{ad}}(N/2)$ in the trivial phase, for $B>B_c$, compared with the distribution of an addition energy separation due to finite-size effect at a lower filling within the lower Hubbard band, $E_{\text{ad}}(N/2)-E_{\text{ad}}(N/2-1)$. One sees that it is highly probable that the gap is much larger than the finite-size energy separations, and thus can be identified in experiments. The critical magnetic field required for the transition to the non-interacting SSH limit varies within a realistic range, as demonstrated in Fig.~\ref{fig:qfdis}b. The critical magnetic field can be reduced if one chooses a smaller hopping amplitude, but this will leave less room for disorder, hence there is a trade off.  A typical conductance spectrum (for $B>B_c$) of the disordered system in Fig.~\ref{fig:qfdis}c shows clearly the high-rising edge-state peak in the non-trivial phase in the middle of the split Hubbard bands of the trivial phase, which can be used for identifying the topological phase transition. We see similar clear signatures in all the 20 random instances we generated for the conductance spectrum. The many-body Zak phase in the non-trivial phase for $B>B_c$ also deviates very little from the ideal value of $\pi$. 

Similar robustness behavior is observed for the cases of pure on-site energy disorder when $\delta E<\vert \Delta t \vert$ and pure hopping amplitude disorder when $\delta t<\vert \Delta t \vert$. The signatures of the topological phases are of course erased for very strong disorders, for example when $\delta t=\delta E = 2 \vert \Delta t \vert$ (see Supplementary Material). In a donor chain in silicon the hopping amplitude oscillates rapidly with the donor separation due to the inter-valley interference in the wavefunction. Even with a positional variation within a silicon unit cell the hopping amplitude can drop to close to zero according to effective mass theory \cite{Le2017}, thus it might be challenging to limit the  hopping amplitude disorder to within the range $[-\vert \Delta t\vert, \vert \Delta t \vert]$. This can be mitigated by fabricating inter-donor side gates depicted as B in Fig.~\ref{fig:exp} for controlling the hopping amplitude. If acceptors are used instead there is no hopping amplitude oscillation owing to the absence of intervalley interference \cite{Zhu2019}, thus the B side gates are not needed, but the A side gates are still required for varying the chemical potential.  

We also investigate the robustness against disorder of the topological phases at half-filling in zero magnetic field, as shown in Fig.~\ref{fig:hf}. Recall that in the large-U limit the properties of the SSHH model can be understood from an effective Heisenberg model of local spins interacting with staggered exchanges, $J_{\pm}=4t_{\pm}^2/U$. The characteristic spin correlation in Fig.~\ref{fig:hf}d, where each dimer in the bulk is strongly correlated while the correlations between the dimers and with the edges are negligible, is preserved if all the odd exchanges are smaller than all the even exchanges. This means the disorder in the hopping amplitude, $\delta t$, should be smaller than $\vert \Delta t \vert$, similar to the quarter-filling case. One interesting difference from the quarter-filling case is that the on-site energy disorder can now be much larger. In the picture of the staggered Heisenberg model one electron is localized at each site, thus changing the on-site energy is akin to changing an energy constant in the Heisenberg Hamiltonian, which does not affect the spin excitation spectrum showing the triplon bands in Fig.~\ref{fig:hf}a. This is true as long as the on-site energy disorder is smaller than the on-site interaction $U$. For stronger variations, where the energy of one site is lower than that of another by more than $U$, double occupancy will be favored, leading to the break down of the Heisenberg picture. When both hopping amplitude and on-site energy disorders are present, we find that the topological phases at half filling are robust when $\delta t<\vert \Delta t \vert$  \emph{and} $\delta t +\delta E < U$. More specifically, the gap above the ground state in Fig.~\ref{fig:hf}a and the characteristic spin-correlation of Fig.~\ref{fig:hf}d remain intact,  and the reduced Zak phase deviates very little from its ideal value. We refer the reader to the Supplementary Material for the numerical results.

Finally, we comment briefly on other possible imperfections. The SSHH model assumes the electrons are phase coherent throughout the length of the chain. A finite chain of 6 sites of dopant atoms in silicon can be made shorter than $50$ nm, while the phase coherence length in STM-fabricated samples at low temperature can be well above $100$ nm, as inferred from weak-localization experiments \cite{Ruess2004}, and in GaAs-based samples the phase coherence length can be as large as a $\mu m$ \cite{Lin2002}. Spin-orbit coupling is not taken into account in our model, but it can be neglected if its energy scale is much smaller than the hopping amplitude. For Si:P, even with the enhancement due to external fields the energy scale of the spin-orbit coupling is of the order of $10^{-6}\,\mu \mathrm{eV}$, as inferred from the spin-flip rate in the region of ms \cite{Weber2018}, which is negligible compared with a hopping amplitude in the meV or sub-meV range. In GaAs spin-orbit coupling may lead to unwanted effect such as spin flip \cite{Scarlino2014} or spin-flip tunneling between the dots \cite{Hofmann2017}, however the energy scale of these effects are just as small. In our calculation of the conductance spectrum we assume an energy-independent tunneling rate, $\Gamma$, between the system and the leads, but in reality electrons tunneling to the upper Hubbard bands have energies much closer to the top of the barrier between the leads and the system, resulting in a larger tunneling rate. The conductance peaks in the upper Hubbard band are therefore should be much higher than those in the lower band. 

In summary, we have investigated the topological phases of a one dimensional Fermi-Hubbard model in the strongly correlated regime. We introduce the concept of the reduced Zak phase, defined based on the reduced density matrix of a subsystem, and show that the topological phases at half-filling can be characterized by this phase. This reduced phase might be useful for studying the topological phases of a subsystem in a larger interacting system, or an open system interacting with the external environment. From a study of entanglement and spin correlation, we demonstrate the bulk-edge correspondence in the half-filled system. At quarter-filling, the model does not exhibit properties of a topological insulator, but it can be transformed to the topological ground state of the non-interacting SSH model by applying a magnetic field. Finally we propose a promising experimental realization with dopant atoms in silicon or quantum dots in GaAs. The scheme is robust against significant disorder in the hopping amplitude and on-site energy.

\section{Methods}
We use the Lanczos algorithm to diagonalize the Hamiltonian in the occupation basis \cite{siro2012}
\begin{equation}
    \ket{1,0,1,\dots,1}_{\uparrow}\otimes\ket{0,1,0,\dots,1}_{\downarrow}, \text{etc.},
\end{equation}
where the first part is for the spin up particles and the second for the spin down ones. A $1$ at position $j$ means the $j$th site is occupied and a $0$ means the reverse. These basis states are then numbered according to the decimal value of its binary string. From this representation the reduced density matrix of the spin-up or spin-down subsystem can be computed in a straight-forward manner. For computing the entanglement entropy between the edges and the rest of the chain, we define the local Hilbert space of each site as $\ket{\varnothing},\ket{\uparrow},\ket{\downarrow},\ket{\uparrow \downarrow}$, that is, a qudit with dimension 4, the entanglement entropy  can then be calculated for the resulting system of qudits \cite{Gu2004}.

\textbf{Data availability}: Data for Nguyen Le et al. “Topological phases of a dimerized Fermi-Hubbard model for semiconductor nano-lattices," https://doi.org/10.5281/zenodo.3346816. The data underlying this work is available without restriction. The Matlab code used in this paper can be downloaded at https://github.com/lehnqt/SSHH.git.

\section{Acknowledgements}We acknowledge financial support from the UK Engineering and Physical Sciences Research Council Grant No. EP/M009564/1 (ADDRFSS) and EPSRC strategic equipment grant no. EP/L02263X/1.

\section{Contributions}
N.H.L. carried out the analytical and numerical calculation. E.G. and N.H.L.  were responsible for the main ideas. N.H.L., A.J.F. and E.G. wrote the manuscript. N.J.C. contributed to the experimental aspect of the proposal. All authors contributed to the discussions and interpretations of the results. 

\section{Competing interests}
The authors declare no competing interests.

\bibliography{SSHHbib}
\end{document}


\title{Supplementary material for Nguyen Le et al ``Topological phases of a dimerized Fermi-Hubbard model for semiconductor nano-lattices''}
\author{Nguyen H. Le}

\affiliation{Advanced Technology Institute and Department of Physics, University of Surrey, Guildford GU2 7XH, United Kingdom}

\author{Andrew J. Fisher}
\affiliation{Department of Physics and Astronomy and London Centre for Nanotechnology,
University College London, Gower Street, London WC1E 6BT, United Kingdom}

\author{Neil J. Curson}
\affiliation{Department of  Electronic and Electrical Engineering and London Centre for Nanotechnology,
University College London, Gower Street, London WC1E 6BT, United Kingdom}

\author{Eran Ginossar}
\affiliation{Advanced Technology Institute and Department of Physics, University of Surrey, Guildford GU2 7XH, United Kingdom}

\maketitle

\subsection{Effect of disorders on the magnetic-field induced topological phases at quarter filling}

\begin{figure*}[h]
\centering
\includegraphics[width=1\textwidth]{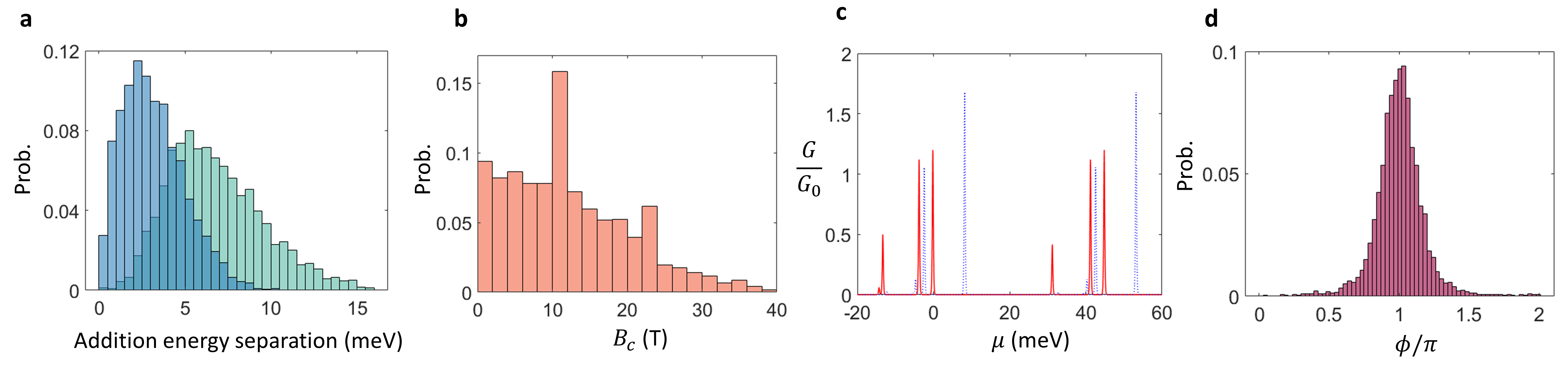}
\caption{\label{fig:qfdis2} As in Fig.~7 of the main text, but with strong disorders, $\delta t=\delta E=2\vert \Delta t \vert$.}
\end{figure*}

In Fig.~7 of the main text we show how the topological phases induced by the magnetic field at quarter filling can still be observed if the disorders in the hopping amplitude (tunnel coupling) and the disorders in the on-site energy are bounded by $\delta t+\delta E < \vert \Delta t \vert$. For the purpose of completeness the case of very strong disorders where it is no longer possible to identify the topological phases is demonstrated in Fig.~\ref{fig:qfdis2}. There is now a significant overlap between the distribution of the quarter-filling gap and that of the finite-size addition energy separation in the band, which means it is likely that the gap cannot be distinguished. The conductance spectrum is now sporadic and no longer shows the high rising edge state peak in the middle of each Hubbard band, and there is large deviation in the many-body Zak phase.

\subsection{Effect of disorders on the topological phases at half filling}
\begin{figure*}[h]
\centering
\includegraphics[width=1\textwidth]{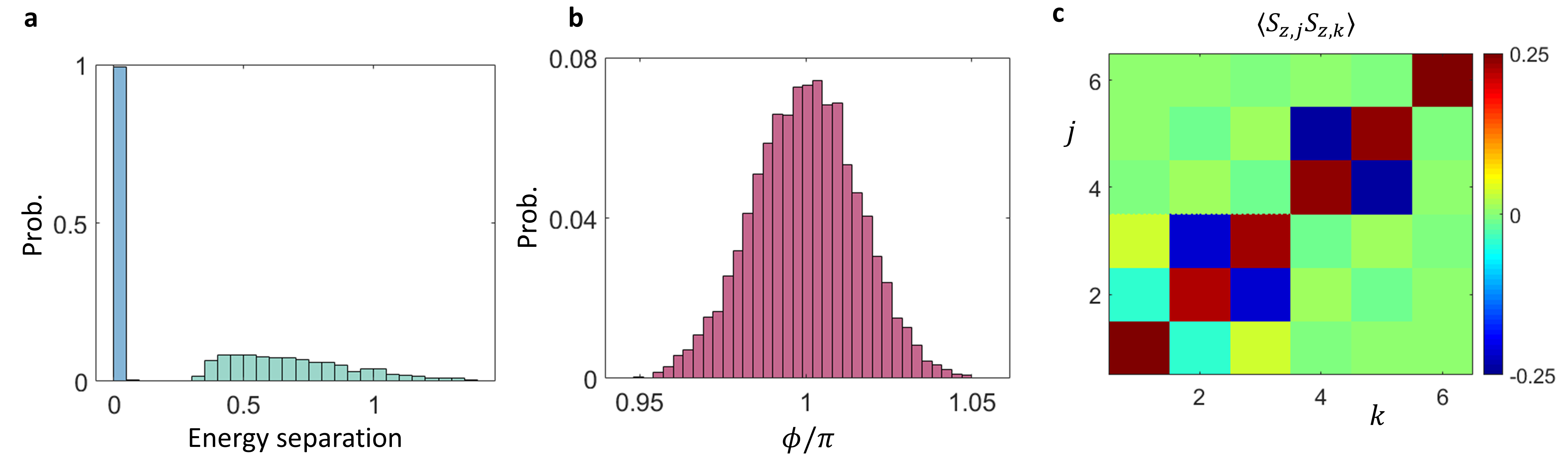}
\caption{\label{fig:hfdis} Robustness of the topological phases at half filling against disorders for a chain with $N=6$ sites. The on-site interaction is $U=10$ as in Fig.~3 of the Main Text, and $\Delta t=0.5$ in the non-trivial phase. The maximum variation in the hopping amplitude disorders is $\delta t= \Delta t = 0.5$, and the maximum variation in the on-site energy disorders is $\delta E = U-\Delta t=9.5$. The distributions are generated from a sample of 5000 random instances. (a) Probability distributions of the energy separation in the degenerate ground state manifold due to disorders (dark blue), and the energy gap between the ground states and the first triplon band (dark green). (b) Probability distribution of the reduced many-body Zak phase, which shows only a small deviation from the ideal value of $\pi$. (c) Spin correlation in the equal mixture of the four degenerate ground states, showing the singlet structure of each dimer in the bulk and the free spins at the edges.}
\end{figure*}

In the Main Text we explained, in the effective picture of the Heisenberg model, how  the topological phases at half filling of the SSHH model in the large-U limit is robust when the disorders in the tunnel coupling is bounded by $\delta t<\vert \Delta t \vert$ and the disorders in the on-site energy satisfies $\delta t+\delta E < U$. In Fig.~\ref{fig:hfdis} we show numerical evidences for the specific case where $\delta t= \vert \Delta t \vert$ and $\delta E=U-\delta t$.

Recall that in the non-trivial phase the ground states are four-fold degenerate due to the spin edge modes, and there is an excitation gap between the ground states and the first triplon band, as shown in Fig.~3a of the Main Text. Figure \ref{fig:hfdis}a shows the distribution of the energy separation between the four ground states caused by disorders and the distribution of the energy gap between them and the first triplon band. There is negligible energy separation in the ground state manifold and the energy gap is always much larger. The reduced many-body Zak phase deviates very little from its ideal value of $\pi$, as seen in Fig.~\ref{fig:hfdis}b. The spin correlation of the equal mixture of the four ground states, which is the thermal state when kT is much smaller than the energy gap, in Fig.~\ref{fig:hfdis}c shows strong correlation between the bulk dimers, but weak correlation between the dimers and between the dimers and the edges. This suggests that each dimer is in a singlet state while the spins at the edges are free, as expected for the non-trivial phase at half filling.

\begin{figure*}[h]
\centering
\includegraphics[width=1\textwidth]{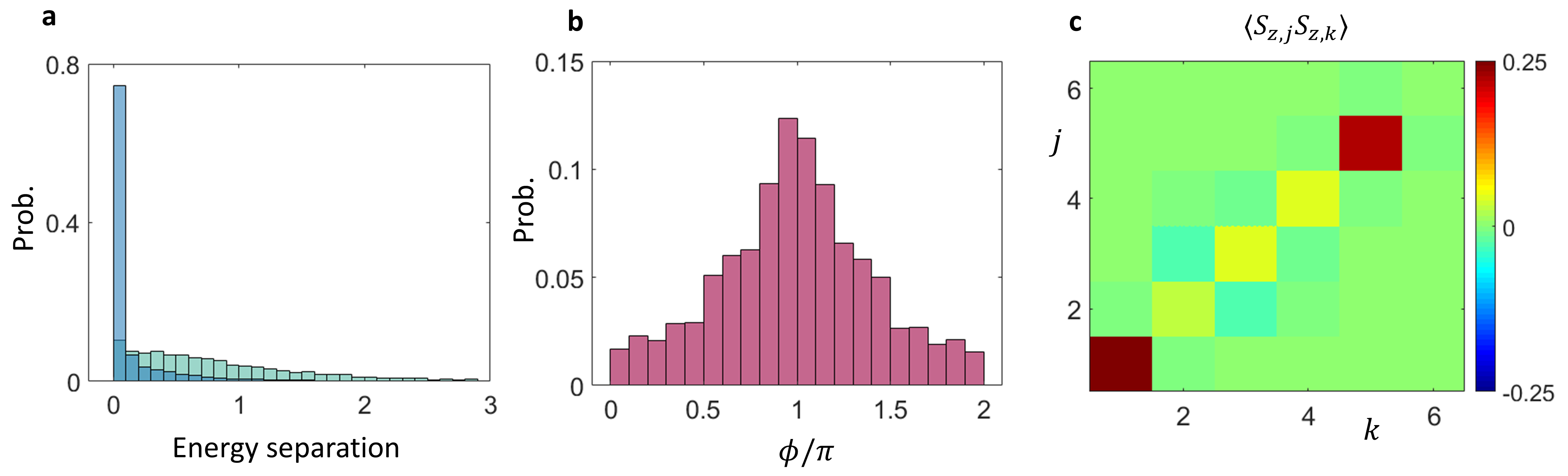}
\caption{\label{fig:hfdis2} As in Fig.~\ref{fig:hfdis}, but with strong disorders, $\delta t=2\Delta t=1$, and $\delta E=2U=20$.}
\end{figure*}

In contrast, when the disorders are very strong, for example $\delta t= 2 \vert \Delta t \vert$ and  $\delta E= 2U$, the characteristics of the topological phases are no longer observed. In Fig.~\ref{fig:hfdis2}  we see a significant overlap of the distribution of the energy separation in the ground state manifold and the energy gap, a very large spread of the reduced many-body Zak phase and the disappearance of the characteristic spin correlation.

\subsection{Charge gap at quarter filling}
We show in Fig.~2 of the Main Text the existence of a gap in the addition energy spectrum at quarter filling and three quarter filling for a chain of $N=12$. We also prove that this gap exists in the thermodynamic limit ($N\rightarrow \infty$) in the fully dimerized regime when the weak hopping amplitude is zero. Here we provide a finite-size scaling analysis of the quarter-filling gap to give further evidence of its existence for the general case when none of the hopping amplitude is zero. 

\begin{figure}[t]
\centering
\includegraphics[width=0.45\textwidth]{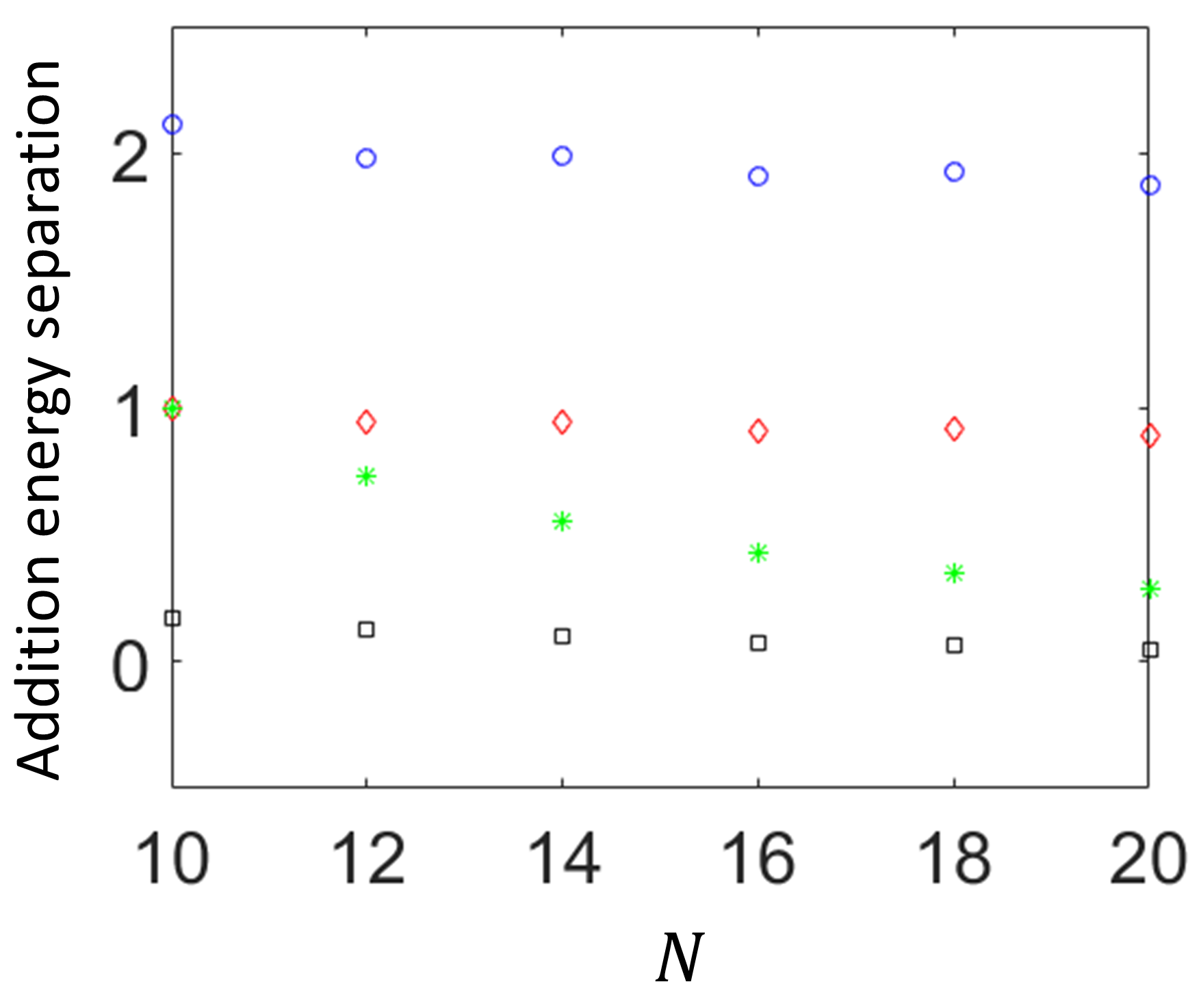}
\caption{\label{fig:qfgap} Finite size scaling analysis of the addition energy gap at quarter filling of the SSHH model, for $U=10$ and $\Delta t=0.5$, as a function of system size $N$. The addition energy gap at quarter filling is shown by the blue circles; its relative value (divided by the initial value at $N=10$) is shown by the red diamonds. An addition energy separation due to finite size effect within the lower Hubbard band, $E_{\text{ad}}(2)-E_{\text{ad}}(1)$, is shown by the black squares and its relative value by the green stars.}
\end{figure}

In Fig.~\ref{fig:qfgap} we show how the addition energy gap at quarter filling and another addition energy separation due to finite size effect within the lower Hubbard band, $E_{\text{ad}}(2)-E_{\text{ad}}(1)$, changes with increasing system size. By looking at their relative changes one notices that the quarter filling gap is almost unchanged while the finite-size energy separation within the band quickly reduces to zero. This shows that the gap at quarter filling remains open while the bands become dense in the thermodynamic limit. A fit of the quarter-filling gap to $a_0+a_1/N+a_2/N^2$ gives $a_0=0.92$, which is the extrapolated gap for $N\rightarrow\infty$. This is close to the hopping amplitude difference, $t_{+}-t_{-}=2\Delta t=0.1$, as expected.

\subsection{Entanglement between the edges and the bulk in the non-interacting limit}

In Fig.~3c in the Main Text we discuss how the entanglement between the edges and the bulk for a half-filled SSHH system should start to drop at $\Delta t = 0$ when crossing from the trivial to the non-trivial phase. This should be the case because this entanglement drop is linked with the formation of the localized edge states which our analysis shows to happen at  $\Delta t > 0$.  Here we provide further numerical proof of this for the non-interacting limit $U=0$.
\begin{figure}[b]
\centering
\includegraphics[width=0.45\textwidth]{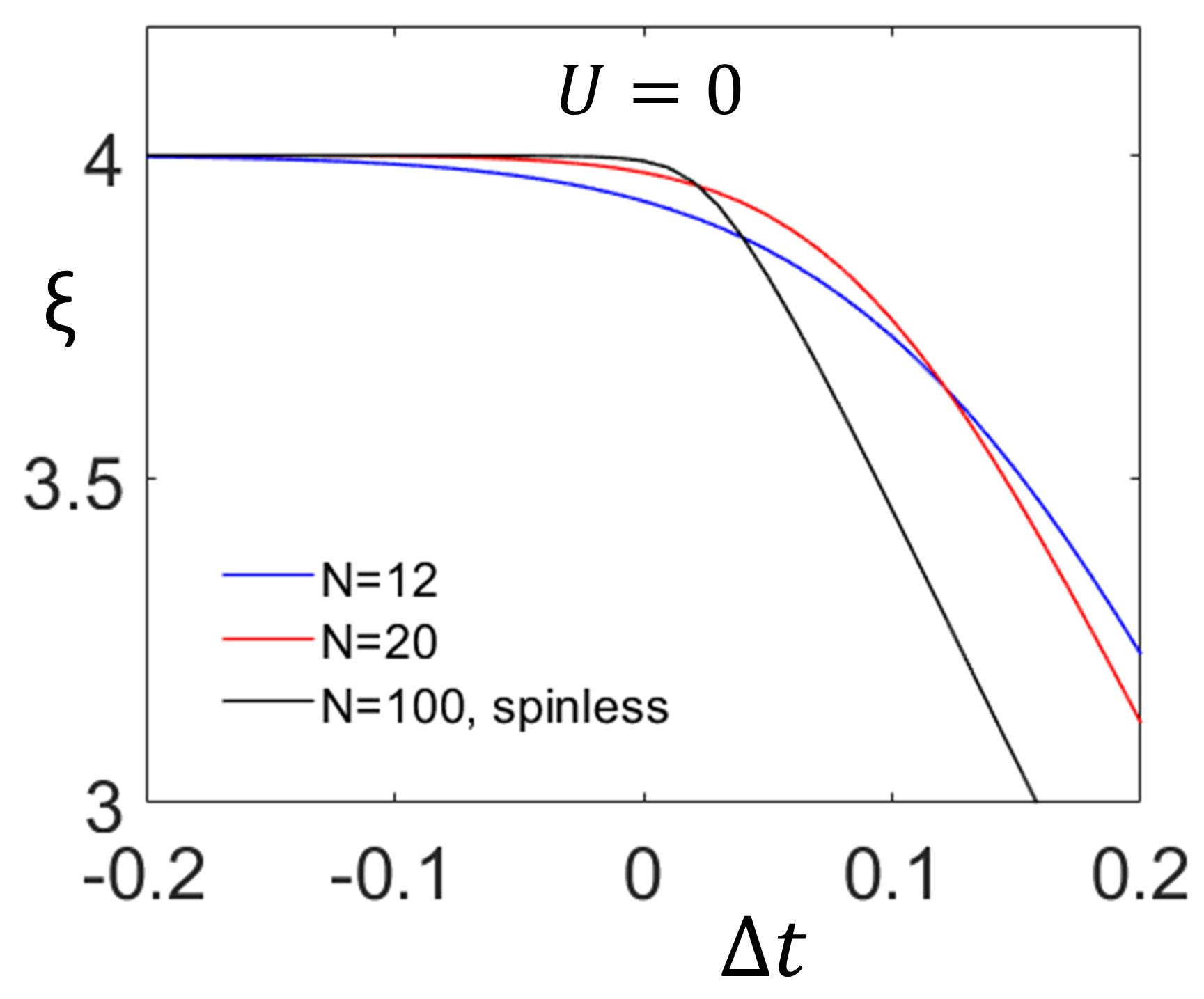}
\caption{\label{fig:sshent} Entanglement between the edges and the bulk for a half-filled SSHH system. The results for the spinful case are shown by the blue curve for a chain with $N=12$ sites and red for $N=20$. The spinless case, where the Pauli exclusion principle is neglected, is shown by the black curve for $N=100$ sites. The entanglement for the spinless cases is normalized to coincide with the entanglement for the spinful case in the trivial phase when $\Delta t<0$.}
\end{figure}

We note that even in the non-interacting limit an exact numerical calculation of the entanglement between the edges and the bulk for a very large chain is not trivial due to the correlation between the non-interacting electrons caused by the Pauli exclusion principle in the many-body wavefunction. The value of the entanglement depends on the definition of the local basis at each site, which includes $\ket{\varnothing},\ket{\uparrow},\ket{\downarrow}$ and  $\ket{\uparrow \downarrow}$ as enforced by the Pauli exclusion principle. An advantage at $U=0$ is that there is no correlation between the spin-up and spin-down electrons and thus we are able to increase the system size to $N=20$ in our exact diagonalization, and find that the entanglement drops when crossing to the non-trivial phase happens closer to the $\Delta t=0$ point as shown in Fig.~\ref{fig:sshent}.

In the spinless case where the Pauli exclusion principle is neglected the many body wavefunction at half filling can be written as a tensor product of the occupied single-particle states, where for each \emph{single-particle} state the local basis is denoted as $\ket{0}$ if the site is not occupied and $\ket{1}$ if the site is occupied. The entanglement is now simply the sum of the entanglements of the single-particle wavefunctions, and it is trivial to do the calculation for a very large number of sites. In Fig.~\ref{fig:sshent} we show that the entanglement between the edges and the bulk at half filling starts to drop at exactly $\Delta t=0$, which is due to the fact that the edge states are localized and the bulk states avoid the edges (see Fig.~1b in the Main Text).
The entanglement does not drop immediately to zero due to the an exponential tail of the edge states penetrating the bulk \cite{asboth2016}. 

The sharp drop at $\Delta t=0$ should survive in the spinful case when the Pauli principle is reintroduced. The reason is that in the non-trivial phase there is little overlap between the electrons at the edges and the electrons in the bulk, so introducing the Pauli principle leads to a change only in the bulk where the electrons overlap. It cannot result in additional correlation between the edges and the bulk. Therefore, the entanglement should drop in the same way as it does in the spinless case.

\bibliography{SSHHbib}